\documentclass[runningheads]{llncs}
\usepackage[T1]{fontenc}
\usepackage{graphicx}%
\usepackage{multirow}%
\usepackage{amsmath,amssymb,amsfonts}%
\usepackage{bm}
\usepackage{array}
\usepackage{mathrsfs}%
\usepackage[title]{appendix}%
\usepackage[dvipsnames]{xcolor}
\usepackage{textcomp}%
\usepackage{manyfoot}%
\usepackage{booktabs}%
\usepackage{algpseudocode}%
\usepackage{listings}%
\usepackage{hyperref}

\begin{document}

\title{Topology, geometry and fabrication of volumetric plain weaves}

\author{Samuel Poincloux\inst{1}\orcidID{0000-0003-4812-2704} \and Lauren Carelli Dreier\inst{2}\orcidID{0000-0003-0558-9099} \and Tian Chen\inst{3,4}\orcidID{0000-0002-9100-7202}}

\institute{Department of Physical Sciences, Aoyama Gakuin University, 5-10-5 Fuchinobe, Sagamihara, 252-5258, Kanagawa, Japan \and Departments of Physics and Mechanical Engineering, University of California, Santa Barbara, CA, USA \and Department of Aeronautics, Imperial College London, London, UK \and Department of Mechanical \& Aerospace Engineering, University of Houston, Houston, TX, USA}

\maketitle

\begin{abstract}
Plain weave is one of the simplest woven structures, formed by two sets of yarns that cross at right angles in an alternating over–under pattern.
While such woven fabrics are usually confined to a 2D surface, here we propose a practical approach to fabricate volumetric plain weaves, entanglements featuring three mutually orthogonal families of yarns threaded through a cubic lattice. 
We develop an analytic framework that converts the crossing pattern topology into continuous yarn centreline geometries, which we then leverage to fabricate volumetric weave specimens. 
Each crossing is encoded by a matrix that defines the offsets of the three yarns at a vertex. These are formulated to preserve the plain weave topology in every plane, where each pair of yarn families interlaces as in a planar plain weave. Eight configurations in four opposite pairs exist, yet all are related by rotations and translations, so that the fully alternating volumetric plain weave has a single topology. We adopt sinusoidal yarn trajectories and show that the crossing matrix defines their phases. We select two solutions: in-phase planar zigzags and out-of-phase circular helices. Both are members of a family of elliptical helices that also admits a free choice of handedness for each yarn. We then derive the conditions under which yarns intersect between vertices. In-phase yarns never intersect, whereas helical yarns of equal handedness intersect at a critical offset of $\sqrt{2}\lambda/16$, with $\lambda$ the wavelength of the cubic lattice. No assignment of handedness avoids these intersections. Finally, we fabricate volumetric plain weaves by 3D printing and manual assembly from cotton cord. The framework provides a direct basis for generating weave geometries for simulation and fabrication.

\end{abstract}

\section{Introduction}

Fabrics are among the most familiar of materials, whose properties depend on both their constituent yarns and the way those yarns interlace. In a woven fabric, the crossing pattern constrains how the yarns move relative to one another, connecting textile structure to physical behaviour. This relationship has motivated research across mathematics~\cite{ghys2011coupe}, computer graphics~\cite{cirio2014yarn}, materials science~\cite{august2020self} and engineering~\cite{grosberg1966mechanical}, where researchers study weaving both as a geometric structure and as a means of designing material response. A planar weave consists of two families of yarns, warp and weft, whose over- and under-crossings repeat in two independent directions. The crossing pattern distinguishes weave types such as plain weave, satin, and twill~\cite{emery1995primary}. In plain weave, each yarn passes alternately over and under successive threads of the other family. The idealized pattern has symmetries that map any yarn onto any other, making plain weave an example of an \emph{isonemal} fabric~\cite{grunbaum1980satins,grunbaum1988isonemal}. Isonemal symmetry expresses the equivalence of the yarns within the pattern and provides a basis for classifying periodic fabrics. A topological description concerns how the yarns interlace, independently of their precise geometry. Methods from knot theory allow textile structures to be compared and classified through their crossing patterns~\cite{grishanov2009topological,fukuda2023classification}. Description of the crossing pattern also lets us ask how yarns hold together as a fabric~\cite{warren2018clothes}. Crossing information can serve a constructive purpose as well. Computational methods have used alternating crossings to generate plain-woven structures over prescribed surfaces and convert them into three-dimensional yarn geometries~\cite{akleman2009cyclic}.

The alternating interlacing of plain weave can also be extended from a surface to a volume. 
3D-woven fabrics, where through-thickness yarns resist delamination between stacked woven layers, are important in engineering composites~\cite{li2022review,wu2020weavecraft}.
However, 3D-woven fabrics are not volumetric weaves, in the sense that their structure cannot be reduced to a volumetric unit cell that can be paved in three directions to construct a 3D entanglement~\cite{green2014mechanical}.
The volumetric weave considered in this paper replaces the two yarn families of a planar fabric with three mutually transverse families threaded through a cubic lattice, producing a periodic entanglement that fills space.
The geometry and topology of tri-periodic entanglements, including weavings and rod  packings, have likewise been studied as mathematical structures~\cite{evans2013periodic,andriamanalina2026untangling}.
Such tri-periodic lattices have also emerged as mechanical metamaterials with compliant and highly stretchable responses~\cite{carton2026woven}, while molecular weaving has been realized in crystalline covalent organic frameworks composed of interlacing helical threads~\cite{liu2016weaving}. 
Recent work has established a constructive lattice-based framework for representing, enumerating, and geometrically realizing volumetric Bravais weaves, including primitive-cubic plain weaves~\cite{volumetric_bravais}. This framework provides a systematic setting in which crossing configurations and their periodic arrangements can be studied. Within such a description, the arrangement at an individual crossing must be compatible with the paths of threads through neighbouring crossings and with the periodicity of the whole structure.

Here, we focus on a periodic cubic weave subject to three conditions. First, each yarn follows a continuous path through successive crossings. Second, its crossings with threads of the other two families alternate, extending the plain-weave pattern into three dimensions. Third, the weave has symmetries that map any yarn onto any other, preserving the isonemal property of the planar pattern. The geometric problem is then to determine continuous yarn trajectories connecting the positions prescribed by the discrete crossing states. For this purpose, we develop an analytic framework that translates the alternating crossing pattern into continuous, periodic yarn centerlines. The construction determines how the crossing states fix the relative phases of the yarn oscillations and identifies realizations in which individual yarns follow planar or helical paths. We then derive geometrical conditions on the yarns' trajectories to avoid spurious crossings that would alter the target topology. Finally, we challenge the framework's validity by constructing volumetric weave samples via additive manufacturing and by manual crafting. This framework makes explicit the connection between the discrete crossing description and the resulting geometry, complementing existing methods for constructing volumetric weaves. It also provides a practical basis for generating weave models for numerical simulation and additive manufacturing, examining yarn intersections, and developing automated checks of weave topology.

\section{Volumetric plain weaves}
We begin with an algebraic description of crossings in planar plain weave, using this case to establish the notation and its geometric interpretation. Note that the description retains all three spatial directions even in the planar case, allowing the same formalism to extend directly to three mutually transverse yarn families on a cubic lattice.

\subsection{Planar weaves and algebraic formalisation of crossing points}

\begin{figure}[ht]
\centering
\includegraphics[width=0.80\textwidth]{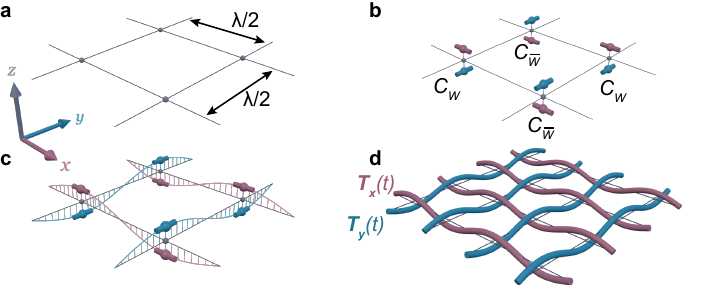}
\caption{\textbf{Planar plain weave.} \textbf{a} Square lattice of crossing sites in the $(x,y)$ plane. \textbf{b} Alternating crossing configurations $C_W$ and $C_{\overline W}$. The weft (pink) and warp (blue) yarns have equal and opposite offsets $\pm a\bm{\hat{z}}$ from the fabric plane. The weft passes over the warp in $C_W$ and under it in $C_{\overline W}$. \textbf{c} Sinusoidal yarn trajectories of Eq.~(\ref{eq_traj2D}) and their translated copies connect the prescribed offsets, with wavelength $\lambda$ and amplitude $a$. \textbf{d} Plain-weave geometry obtained by periodically repeating the yarn trajectories.}\label{fig_2Dweave}
\end{figure}

We consider a 3-dimensional space equipped with a direct orthonormal coordinate system $(x,y,z)$ with corresponding unit directional vectors $(\bm{\hat{x}},\bm{\hat{y}},\bm{\hat{z}})$. 
The frame is oriented such that the weft yarns run along the $\bm{\hat{x}}$ direction, and the warp yarns along the $\bm{\hat{y}}$ direction. 
Warp and weft yarns cross at each vertex of a lattice in the $(\bm{\hat{x}},\bm{\hat{y}})$ plane with periodicity $\lambda_x/2$ in the $\bm{\hat{x}}$ direction and $\lambda_y/2$ in the $\bm{\hat{y}}$ direction (Fig.~\ref{fig_2Dweave}a). 
At each crossing, the two yarns are displaced by a distance $a$ on opposite sides of the fabric plane. The yarn passing over has offset $+a\bm{\hat{z}}$, while the yarn passing under has offset $-a\bm{\hat{z}}$ (see Fig.~\ref{fig_2Dweave}b).
We represent each crossing by a $3\times3$ crossing matrix $C=[c_{ij}]$, with rows and columns indexed by $x,y,z$ and entries equal to $-1$, $0$, or $+1$. 
Each row of this matrix represents one yarn direction, while the columns represent the normalised offset in the corresponding direction. 
In the 2D planar case, the fabric consists only of weft ($x$) and warp ($y$) yarns, and the offsets are only in the normal $z$ direction. 
The crossing matrix is then reduced to only its $c_{xz}=\pm1$ and $c_{yz}=\pm1$ elements, all other elements being $0$. 
At each crossing, the warp and weft offset vectors are equal and opposite, so $c_{xz}+c_{yz}=0$. The two possible configurations, weft over warp ($W$) and warp over weft ($\overline W$), are represented by the crossing matrices $C_W$ and $C_{\overline W}$, respectively: 
\begin{align}
& C_W=\left[{  \begin{array}{rrr} 0 & \quad0 & 1 \\ 0 & 0 & -1 \\ 0 & 0 & 0 \end{array}}\right] & C_{\overline W}=\left[{  \begin{array}{rrr} 0 & \quad0 & -1 \\ 0 & 0 & 1 \\ 0 & 0 & 0 \end{array}}\right].
\end{align}
Each crossing is therefore fully specified by the sign $z_\bot=c_{xz}$, with $z_\bot=+1$ for weft over warp ($W$) and $z_\bot=-1$ for warp over weft ($\overline W$). 
The crossing pattern of a weave is thereby defined by the collection of $z_\bot=\pm1$ values assigned to its lattice sites. 

In plain weave, $z_\bot$ changes sign between neighbouring lattice sites along both yarn directions, giving the repeating $2\times2$ crossing pattern
\[
\begin{bmatrix}
W & \overline W\\
\overline W & W
\end{bmatrix}.
\]
Representing $z_\bot=+1$ and $-1$ by two colours gives the traditional bicolour grid and its checkerboard arrangement (see Fig.~\ref{fig_2Dweave}b). The fabric is thus biperiodic, repeating every two crossings along both the $x$ and $y$ directions, so a single repeating unit cell specifies its entire crossing pattern. Exchanging $W$ and $\overline W$ gives an equivalent plain weave, obtained by translating the pattern by half a wavelength along either direction.

Having specified the crossing pattern, we now construct yarn trajectories that accommodate the finite diameter of physical yarns. The yarns must undulate out of the fabric plane to pass alternately over and under one another without intersecting. We model this undulation using sinusoidal trajectories that pass through the offsets prescribed by the crossing matrix \(C\).
Taking \(\lambda_x=\lambda_y=\lambda\), we consider a \(2\times2\) arrangement of crossings at the corners of a square of side length \(\lambda/2\) (see Fig.~\ref{fig_2Dweave}), with one corner at the origin. The configuration there is \(C_W\) or \(C_{\overline W}\), corresponding to \(z_\bot=\pm1\). Since the crossing configurations alternate along both directions, the two adjacent corners have the opposite value, \(-z_\bot\), while the diagonally opposite corner has the same value, \(z_\bot\).
For the weft and warp yarns passing through the origin, we propose the parametric trajectories shown in Fig.~\ref{fig_2Dweave}c:

\begin{align}
\bm{T_x}(t)&=\left|{\begin{array}{l} x=t\\ y=0\\ z=a\cos(2\pi t /\lambda +\Phi_{xz}) \end{array} }\right. &
\bm{T_y}(t)&=\left|{\begin{array}{l} x=0\\ y=t \\ z=a\cos(2\pi t /\lambda +\Phi_{yz}) \end{array} }\right. \label{eq_traj2D}
\end{align}
Each yarn extends along its own direction with the parameter $t$ in the plane of the fabric and oscillates normal to it with wavelength $\lambda$ and amplitude $a$. The wavelength is twice the crossing spacing, so that consecutive crossings fall on a maximum and a minimum of the oscillation; the phases $\Phi_{xz}$ and $\Phi_{yz}$ are fixed by the configuration at the origin. To express this condition, we gather the trajectories of the yarns through the origin in the yarn trajectory matrix:
\begin{equation}
\underline{\underline{T}}(t)=\left[{\begin{array}{l} \bm{T_x}(t)\\ \bm{T_y}(t)\\ \bm{0} \end{array} }\right]=\left[{\begin{array}{ccc} t \quad & 0 \quad & a\cos(2\pi t/\lambda+\Phi_{xz}) \\ 0 \quad & t \quad & a\cos(2\pi t/\lambda+\Phi_{yz}) \\ 0 \quad & 0 \quad & 0 \end{array}}\right],
\label{eq_yarnmatrix2D}
\end{equation}
whose rows are indexed like those of the crossing matrix, the third row being empty in the absence of a yarn along $z$. At the origin, the yarns must pass through the offsets prescribed by the crossing matrix, $\underline{\underline{T}}(0)=aC$, which gives the two equations
\begin{equation}
\cos\Phi_{xz}=c_{xz}=z_\bot,\qquad \cos\Phi_{yz}=c_{yz}=-z_\bot.
\end{equation}
The phase conditions give
\[
\Phi_{xz}=\frac{\pi}{2}(1-z_\bot),
\qquad
\Phi_{yz}=\frac{\pi}{2}(1+z_\bot)
\pmod{2\pi},
\]
so the warp and weft oscillate in antiphase. At successive crossings, $t=n\lambda/2$, the identity
\[
\cos(n\pi+\Phi_{iz})=(-1)^n\cos\Phi_{iz},\, \,i\in\{x,y\}
\]
ensures that the yarn offsets alternate as prescribed by the crossing pattern. For fixed amplitude $a$ and wavelength $\lambda$, these phase conditions uniquely determine the trajectories within the chosen sinusoidal form.

The neighbouring yarns have the opposite crossing states. The weft on $y=\lambda/2$ follows
\[
\bm{T_x}(t)
+\frac{\lambda}{2}\bm{\hat{x}}
+\frac{\lambda}{2}\bm{\hat{y}},
\]
and the warp on $x=\lambda/2$ follows
\[
\bm{T_y}(t)
+\frac{\lambda}{2}\bm{\hat{y}}
+\frac{\lambda}{2}\bm{\hat{x}}.
\]
Each trajectory is translated by half a wavelength along its own direction and by half a wavelength onto the adjacent lattice line. The translation along the yarn reverses its normal displacement at the lattice sites. Together with the two yarns through the origin, these trajectories complete the geometric construction of the unit cell. Paving this unit cell in the weft and warp directions reconstructs a plain weave, as shown in Fig.~\ref{fig_2Dweave}d.
 
\subsection{Generalization to volumetric weaves}
\label{generalization}
\begin{figure}[ht]
\centering
\includegraphics[width=0.66\textwidth]{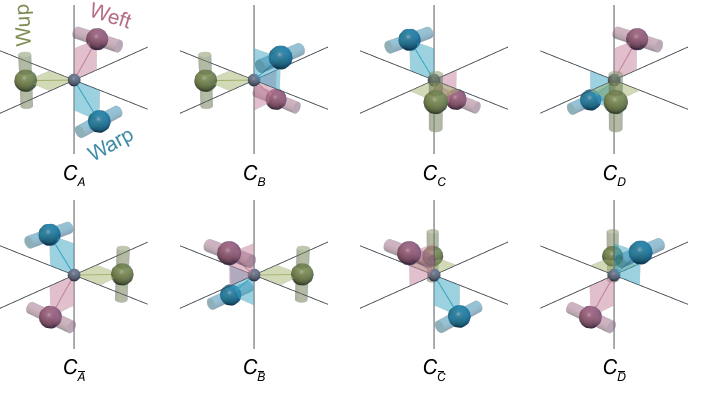}
\caption{\textbf{The eight crossing configurations of a volumetric weave.} At a vertex, the weft (pink, along $x$), the warp (blue, along $y$) and the wup (green, along $z$) are each offset by $\pm a$ in the two directions perpendicular to them. The sphere marks the offset point of each yarn; along each direction, the offsets of the two yarns perpendicular to it are opposite. The top row shows $C_A$ to $C_D$ of Eq.~(\ref{eq_configs}), the bottom row their opposites $C_{\overline X}=-C_X$, in which every offset is reversed.
}\label{fig_3Dcross}
\end{figure}

We now consider volumetric weaves, which we define as 3-periodic entanglements whose entanglement patterns resemble weaves when sliced perpendicular to each periodicity direction. 
To define such weaves, we follow the same framework developed for planar weaves in the previous section. 
We keep a 3-dimensional space equipped with a direct orthonormal coordinate system $(x,y,z)$, with weft yarns along the $x$ direction and warp ones along the $y$ direction. 
We introduce the \textit{wup} yarns as a third family of yarns that run upward along the $z$ direction. 
We can then construct a cubic lattice with periodicity $(\lambda_x/2,\lambda_y/2,\lambda_z/2)$, such that one weft, one warp, and one wup yarn cross at each vertex. 
By extending the definition of offset vectors introduced for planar weaves to the volumetric case, a yarn at a crossing point is now offset in two perpendicular directions (see Fig.~\ref{fig_3Dcross}). 
For instance, a wup yarn is offset by $\pm a\bm{\hat{x}}$ and $\pm a\bm{\hat{y}}$, a weft yarn by $\pm a\bm{\hat{y}}$ and $\pm a\bm{\hat{z}}$ and a warp yarn by $\pm a\bm{\hat{x}}$ and $\pm a\bm{\hat{z}}$. 
To preserve the over--under arrangement, the two yarns perpendicular to a given coordinate direction are assigned equal and opposite offsets along that direction. For example, if the wup yarn is offset by $+a\bm{\hat{x}}$, the warp yarn must be offset by $-a\bm{\hat{x}}$. 
The offsets at each crossing therefore sum to zero in each of the three coordinate directions. The offset and cancellation conditions lead to eight possible crossing configurations, grouped into four pairs in which every over--under relation is reversed. To identify these configurations, we collect the yarn offsets in the crossing matrix $C=[c_{ij}]$ and express the cancellation condition in each coordinate direction. 

The crossing matrix $C$ retains the $3\times3$ form and indexing introduced for planar weaves. In the planar case, only $c_{xz}$ and $c_{yz}$ are non-zero, recording the opposing weft and warp offsets normal to the fabric. In the volumetric case, each yarn family has offsets in two perpendicular directions. The matrix therefore has six non-zero off-diagonal entries, each equal to $\pm1$, while its diagonal entries remain zero, $c_{ii}=0$. The offsets must still cancel in each direction, so each column sums to zero, $\sum_i c_{ij}=0$. These three conditions link the six non-zero entries, leaving three independent sign choices and therefore $2^{6-3}=8$ possible crossing configurations. The corresponding matrices are listed below (see Fig.~\ref{fig_3Dcross}).

\begin{align}
& C_A=\left[{  \begin{array}{rrr} 0 & 1 & 1 \\ 1 & 0 & -1 \\ -1 & -1 & 0 \end{array}}\right] & C_{\overline A}=\left[{  \begin{array}{rrr} 0 & -1 & -1 \\ -1 & 0 & 1 \\ 1 & 1 & 0 \end{array}}\right] \nonumber \\ \nonumber\\
& C_B=\left[{  \begin{array}{rrr} 0 & 1 & -1 \\ 1 & 0 & 1 \\ -1 & -1 & 0 \end{array}}\right]  & C_{\overline B}=\left[{  \begin{array}{rrr} 0 & -1 & 1 \\ -1 & 0 & -1 \\ 1 & 1 & 0 \end{array}}\right] \nonumber  \\ \label{eq_configs} \\
& C_C=\left[{  \begin{array}{rrr} 0 & 1 & -1 \\ -1 & 0 & 1 \\ 1 & -1 & 0 \end{array}}\right] & C_{\overline C}=\left[{  \begin{array}{rrr} 0 & -1 & 1 \\ 1 & 0 & -1 \\ -1 & 1 & 0 \end{array}}\right] \nonumber \\ \nonumber\\
& C_D=\left[{  \begin{array}{rrr} 0 & 1 & 1 \\ -1 & 0 & -1 \\ 1 & -1 & 0 \end{array}}\right]  & C_{\overline D}=\left[{  \begin{array}{rrr} 0 & -1 & -1 \\ 1 & 0 & 1 \\ -1 & 1 & 0 \end{array}}\right] \nonumber
\end{align}
These eight configurations form four pairs, with $C_{\overline X}=-C_X$ for $X\in\{A,B,C,D\}$. Within each pair, all over--under relations are reversed, as in the $W$ and $\overline W$ configurations of planar weave. Along each coordinate direction, the two yarns perpendicular to it are assigned offsets $+a$ and $-a$. Their normalised offsets are therefore $+1$ and $-1$, giving equal and opposite entries in the corresponding column. Choosing one sign in each column fixes the other entry, so three signs specify the entire crossing configuration. We choose $x_\bot=c_{yx}$, $y_\bot=c_{xy}$ and $z_\bot=c_{xz}$, each equal to $\pm1$, giving
\begin{equation}
C_{(x_\bot,y_\bot,z_\bot)}=\left[{  \begin{array}{rrr} 0 & y_\bot & z_\bot \\ x_\bot & 0 & -z_\bot \\ -x_\bot & -y_\bot & 0 \end{array}}\right]
\end{equation}
Reversing all three signs gives the opposite configuration. For example, $C_A=C_{(1,1,1)} = -C_{(-1,-1,-1)}=-C_{\overline{A}}$.

We define a volumetric equivalent of the 2D alternating plain weave as an alternating tiling of opposite-pair configurations in all directions. 
A cut perpendicular to any of the three directions must give a plain weave pattern, alternating between warp/weft, warp/wup or weft/wup, depending on the direction of the cut (see Fig.~\ref{fig_3Dunit}d).
For example, a volumetric plain weave can be constructed by alternating $A$ and $\overline A$ configurations in all 3 directions. 
The basic unit cell is then a $2\times2\times2$ cube of dimensions $(\lambda_x/2,\lambda_y/2,\lambda_z/2)$, with configurations
\[
\left[
\begin{bmatrix}
X & \overline X\\
\overline X & X
\end{bmatrix},
\quad
\begin{bmatrix}
\overline X & X\\
X & \overline X
\end{bmatrix}
\right],
\qquad X\in\{A,B,C,D\}.
\]
This cell is tiled along the three coordinate directions (see Fig.~\ref{fig_3Dunit}a). 
A notable implication is that reversing every crossing is equivalent to translating the weave by half a wavelength along any lattice direction, just as in planar plain weave. This translation exchanges $X$ and $\overline X$ at every vertex, so either member of a pair can occupy the origin without changing the periodic weave.

The four pairs $X/\overline X$, with $X\in\{A,B,C,D\}$, initially provide four choices for constructing an alternating volumetric plain weave. When the three yarn families are identical and $\lambda_x=\lambda_y=\lambda_z$, the cubic lattice is preserved by rotations through $\pi$ about its coordinate axes. Let $R_i(\pi)$ denote the three-dimensional rotation matrix about axis $\bm{\hat{i}}$, with $i\in\{x,y,z\}$. The crossing configurations satisfy
\begin{align}
    C_A\times R_x(\pi) & = C_{\overline D} = - C_D \nonumber \\
    C_A\times R_y(\pi) & = C_{C}   \\
    C_A\times R_z(\pi) & = C_{\overline B} = - C_B \nonumber 
\end{align}
Each rotation preserves the relation between a configuration and its opposite, since $(-C_A)R_i(\pi)=-C_A R_i(\pi)$. The rotations also preserve the cubic lattice and the alternation between neighbouring crossings. The relations above therefore map the entire $A/\overline A$ tiling onto those formed from the other three pairs. Combined with translation by half a wavelength exchanging the two members within each pair, these rotations and translations relate all four choices, establishing one 3-periodic topology within the fully alternating cubic construction considered here. In the following section, we construct continuous yarn trajectories for this crossing pattern.

\subsection{Yarn trajectories}\label{sec_traj}

The alternating crossing pattern developed in Section~\ref{generalization} prescribes the plain weave topology via the yarn offsets at each lattice vertex. We now construct continuous yarn trajectories that pass through these prescribed positions. The crossing conditions constrain the trajectories at the vertices while leaving freedom in the paths between them.

\begin{figure}[ht]
\centering
\includegraphics[width=\textwidth]{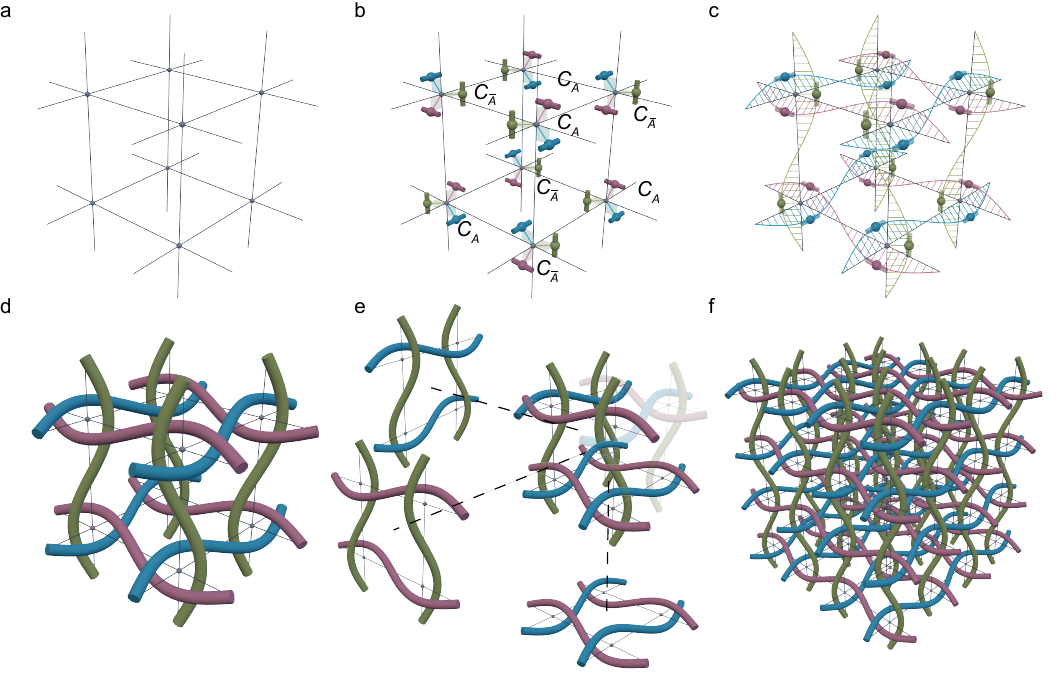}
\caption{\textbf{Volumetric plain weave.} \textbf{a} The $2\times2\times2$ unit cell, with edge length of $\lambda/2$. \textbf{b} The spheres mark the offsets of the weft (pink), the warp (blue) and the wup (green); neighbouring vertices carry opposite configurations $C_X$ and $C_{\overline X}$. \textbf{c} The same cell with the yarns following the in-phase trajectories of Section~\ref{sec_traj}, each passing through its offset points. \textbf{d} A complete unit. \textbf{e} Sections of the cell perpendicular to $x$, $y$ and $z$: each is a planar plain weave of the two yarn families it contains, warp--wup, weft--wup and weft--warp respectively. \textbf{f} A larger weave obtained by tiling the cell.}\label{fig_3Dunit}
\end{figure}

We consider a $2 \times 2 \times 2$ cubic unit cell of side length $\lambda/2$, with a crossing configuration at each vertex (see Fig.~\ref{fig_3Dunit}b). 
The orthonormal frame $(x,y,z)$ places one vertex at $(0,0,0)$ and its diagonally opposite vertex at $(\lambda/2,\lambda/2,\lambda/2)$. 
Choosing the configuration $C_{(x_\bot,y_\bot,z_\bot)}$ at the origin fixes the configurations at the remaining vertices through the fully alternating construction.

We begin with the three yarns that form the crossing around the lattice vertex $(0,0,0)$. For the 2D plain weave case, we describe each yarn by a three-dimensional position vector whose two transverse components vary sinusoidally along its principal direction. We choose a wavelength $\lambda$, twice the spacing between neighbouring vertices, so that the transverse offsets reverse sign at successive crossings.
We propose the following parametric functions $(\bm{T_x}(t), \bm{T_y}(t), \bm{T_z}(t))$ for respectively the weft, warp and wup yarns: 
\begin{align}
\bm{T_x}(t)&=\left|{\begin{array}{l} x=t\\ y=a\cos(2\pi t /\lambda +\Phi_{xy})\\ z=a\cos(2\pi t /\lambda +\Phi_{xz}) \end{array} }\right. \nonumber \\
\bm{T_y}(t)&=\left|{\begin{array}{l} x=a\cos(2\pi t /\lambda +\Phi_{yx})\\ y=t \\ z=a\cos(2\pi t /\lambda +\Phi_{yz}) \end{array} }\right.\nonumber \\
\bm{T_z}(t)&=\left|{\begin{array}{l} x=a\cos(2\pi t /\lambda +\Phi_{zx})\\ y=a\cos(2\pi t /\lambda +\Phi_{zy})  \\ z=t \end{array} }\right.
\end{align}
Each transverse component has amplitude $a$, equal to the prescribed offset magnitude. We determine its phase by matching the trajectory to the crossing configuration. The parameter $t$ specifies position along each yarn's principal axis. For the weft, $x=t$, while the $y$- and $z$-coordinates oscillate sinusoidally. The warp and wup follow the same construction along $y$ and $z$, respectively. The phase $\Phi_{ij}$ determines the starting position of the oscillation in direction $j$ for the yarn running along direction $i$. 

To connect the trajectories to the crossing matrix, we collect the three yarn paths into a single matrix, extending the yarn trajectory matrix introduced for planar weaves (see Eq.~(\ref{eq_yarnmatrix2D})).
\begin{equation}
\underline{\underline{T}}(t)=\left[{\begin{array}{l} \bm{T_x}(t)\\ \bm{T_y}(t)\\ \bm{T_z}(t) \end{array} }\right] = \left[{  \begin{array}{rrr} 0 \quad & \bm{T_x}\cdot \bm{\hat{y}} \quad & \bm{T_x}\cdot \hat{\bm{z}} \\ \bm{T_y}\cdot \bm{\hat{x}} \quad & 0 \quad & \bm{T_y}\cdot \bm{\hat{z}} \\ \bm{T_z}\cdot \hat{\bm{x}} \quad & \bm{T_z}\cdot \bm{\hat{y}} \quad & 0 \end{array}}\right] + t\underline{\underline{I}}
\end{equation}
Each row contains the coordinates of one yarn, and the columns correspond to the $x$, $y$ and $z$ directions. The off-diagonal entries describe the transverse oscillations, expressed using the unit vectors $\bm{\hat{x}}$, $\bm{\hat{y}}$ and $\bm{\hat{z}}$. The term $t\underline{\underline{I}}$, where $\underline{\underline{I}}$ is the identity matrix, supplies each yarn's position along its principal axis. At $t=0$, the transverse offsets must match those prescribed by the crossing configuration at the origin, giving

\begin{equation}
    \underline{\underline{T}}(t=0)=a C_{(x_\bot,y_\bot,z_\bot)}.
\end{equation}

The next step is to choose the phases so that the curves pass through the offsets prescribed by the crossing matrix. At $t=0$, each transverse coordinate in Eq.~(8) reduces to $a\cos\Phi_{ij}$, while the corresponding entry in Eq.~(10) prescribes an offset of $ac_{ij}$. Cancelling the common amplitude $a$ gives six phase conditions, one for each transverse coordinate.
\begin{equation}
\left[{  \begin{array}{ccc} 0 & \cos\Phi_{xy} & \cos\Phi_{xz} \\ \cos\Phi_{yx} & 0 & \cos\Phi_{yz} \\ \cos\Phi_{zx} & \cos\Phi_{zy} & 0 \end{array}}\right]  = \left[{  \begin{array}{rrr} 0 & y_\bot & z_\bot \\ x_\bot & 0 & -z_\bot \\ -x_\bot & -y_\bot & 0 \end{array}}\right].
\end{equation}

Equivalently, using the entries $c_{ij}$ of the crossing matrix, these conditions can be written as
\begin{equation}
\forall (i,j), \, i\neq j, \, \cos\Phi_{ij}=c_{ij}.
\label{eq_cross}
\end{equation}
The yarns must also verify the alternating crossings $\underline{\underline{T}}(t=n \lambda/2)-\frac{n \lambda}{2}\underline{\underline{I}}=(-1)^n a\,C_{(x_\bot,y_\bot,z_\bot)}, n\in\mathbb{Z}$. 
However, this is directly enforced by the periodicity of the functions, as $\cos(2\pi t /\lambda+ \Phi_{ij})|_{t=n\lambda/2}= \cos(n\pi + \Phi_{ij})=(-1)^n\cos(\Phi_{ij})$. 
We then develop two solutions that satisfy the prescribed crossing conditions, which we call in-phase and out-of-phase.

\subsubsection{In-phase trajectories} 

To find a first solution of Eq.~\ref{eq_cross}, since each non-zero entry $c_{ij}$ is either $+1$ or $-1$, the phase conditions are satisfied by choosing $\Phi_{ij}=0$ or $\pi$, respectively, modulo $2\pi$. These choices define the in-phase solution and can be expressed as $\Phi_{ij}=(1-c_{ij})\pi/2\, [2\pi]$, or written in full,
\begin{align}
\Phi_{xy}&=(1-y_\bot)\frac{\pi}{2}\, [2\pi], &
\Phi_{yx}&=(1-x_\bot)\frac{\pi}{2}\, [2\pi], &
\Phi_{zx}&=(1+x_\bot)\frac{\pi}{2}\, [2\pi] \nonumber \\
\Phi_{xz}&=(1-z_\bot)\frac{\pi}{2}\, [2\pi], &
\Phi_{yz}&=(1+z_\bot)\frac{\pi}{2}\, [2\pi], &
\Phi_{zy}&=(1+y_\bot)\frac{\pi}{2}\, [2\pi], 
\label{eq_inphase}
\end{align}
For these phases, each transverse oscillation is the same cosine multiplied by its prescribed sign. The two transverse coordinates of each yarn are therefore equal or opposite throughout its path, confining the yarn to a fixed oblique plane. Sections perpendicular to $x$, $y$ and $z$ show planar plain-weave patterns between the warp--wup, weft--wup, and weft--warp yarns, respectively (see Fig.~\ref{fig_3Dunit}d). A tiled example of an in-phase volumetric plain weave is shown in Fig.~\ref{fig_3Dunit}e.

\subsubsection{Out-of-phase trajectories}
To construct the second solution, we introduce a quarter-period lag between each yarn's two transverse oscillations. Maintaining the prescribed offsets at the crossings then requires an oscillation amplitude of $\sqrt{2}\,a$, such that $a\rightarrow\sqrt{2}a$.

For a yarn along direction $i$, let $(j,j')$ be its two transverse directions taken in the cyclic order of $(x,y,z)$, that is $(y,z)$ for the weft, $(z,x)$ for the warp and $(x,y)$ for the wup, so that $\Phi_{ij}$ and $\Phi_{ij'}$ are the phases of its two transverse oscillations. 
For this specific solution, we take the phases as odd multiples of $\pi/4$, with a quarter-period lag between the two transverse oscillations of each yarn, $\Phi_{ij'}-\Phi_{ij}=-\pi/2$. 
The yarn is then a helix: its transverse coordinates are $\sqrt2\,a\cos(2\pi t/\lambda+\Phi_{ij})$ and $\sqrt2\,a\sin(2\pi t/\lambda+\Phi_{ij})$, and it turns from direction $j$ toward direction $j'$ as it advances, which makes it right-handed. 
 
\begin{figure}[!ht]
\centering
\includegraphics[width=\textwidth]{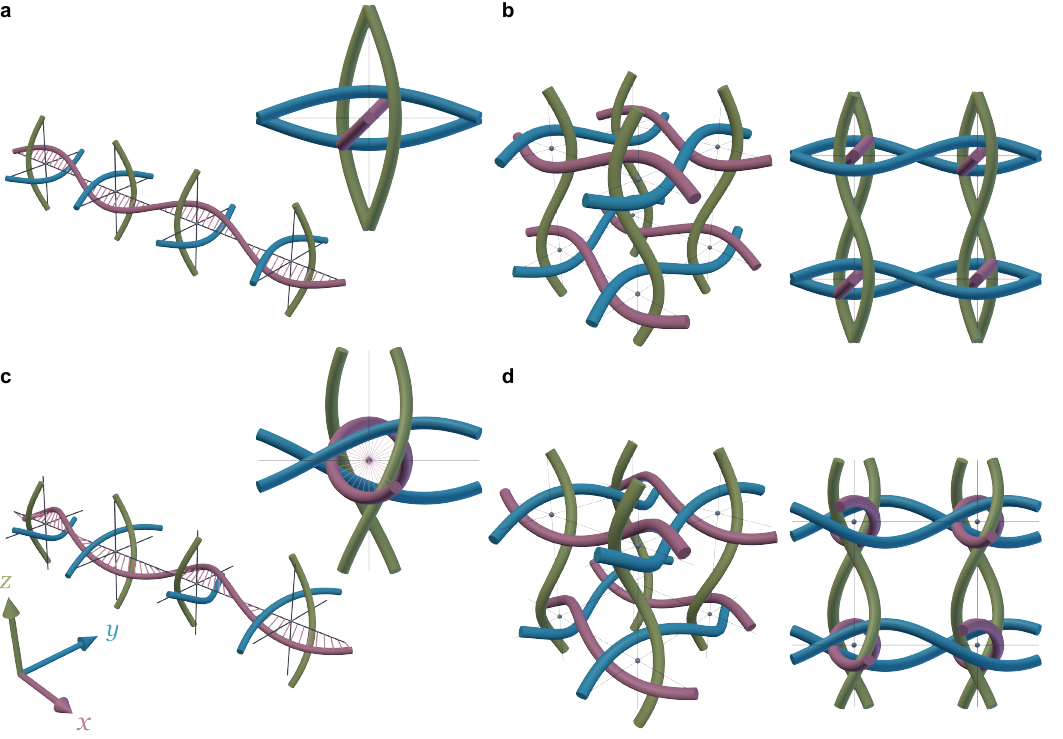}
\caption{\textbf{In-phase and out-of-phase trajectories.} \textbf{a} In-phase weft (pink) through consecutive vertices. The offset keeps a fixed direction and the yarn is a planar zigzag. \textbf{b} Out-of-phase weft through the same offset points: the offset turns about the lattice line and the yarn is a helix of radius $\sqrt2\,a$. \textbf{c}, \textbf{d} The $2\times2\times2$ unit cell with in-phase and out-of-phase yarns.}\label{fig_helix}
\end{figure}

The conditions at the vertex become
\begin{align}
\sqrt{2}\cos\Phi_{ij}&=c_{ij}, & \sqrt{2}\sin\Phi_{ij}&=c_{ij'},
\end{align}
which identify $\Phi_{ij}$ with the polar angle, in the transverse plane $(j,j')$ of the yarn, of the corner $(c_{ij},c_{ij'})$ through which it passes at the vertex: 
\begin{equation}
\Phi_{ij}=c_{ij'}(2-c_{ij})\frac{\pi}{4}\, [2\pi],\qquad \Phi_{ij'}=\Phi_{ij}-\frac{\pi}{2}. \label{eq_helix}
\end{equation}
 In explicit form, we get: 
 \begin{align}
\Phi_{xy}&=z_\bot(2-y_\bot)\frac{\pi}{4}\, [2\pi], & \Phi_{xz}&=\Phi_{xy}-\frac{\pi}{2}, \nonumber \\
\Phi_{yz}&=x_\bot(2+z_\bot)\frac{\pi}{4}\, [2\pi], & \Phi_{yx}&=\Phi_{yz}-\frac{\pi}{2}, \nonumber \\
\Phi_{zx}&=-y_\bot(2+x_\bot)\frac{\pi}{4}\, [2\pi], & \Phi_{zy}&=\Phi_{zx}-\frac{\pi}{2}. 
\end{align}
The two solutions share the same topology but differ significantly in their yarn geometry (see Fig.~\ref{fig_helix}). We anticipate that these differences could have implications for manufacturing and mechanical behaviour~\cite{bertoldi2017flexible,wang2021structured,gonzalez2025pulling,moestopo2020pushing}.

\subsubsection{Generalized trajectories}
The two trajectories constructed above can also be viewed as endpoints of a continuous interpolation from a planar path to a circular helix. We thus move toward a more general description of a family of trajectories obtainable between these two cases. Previously, we went from a planar trajectory to a circular helix by shifting the phases of the two transverse components by equal and opposite amounts of $\pi/4$. Viewed end-on, these trajectories appear as a diagonal line at $45^\circ$ to the coordinate axes and a circle, respectively (see Fig.~\ref{fig_helix}a,c).

Here, we note that the phase shifts can also take any intermediate value between zero and $\pi/4$, giving elliptical paths when viewed end-on. These naturally interpolate between the diagonal line and the circle. As the phase shifts vary, we adjust the amplitude to preserve the offsets prescribed by the crossing matrix. Throughout this interpolation, the semi-axis along the diagonal through the prescribed offsets remains fixed at $\sqrt2\,a$, while the perpendicular semi-axis, $b$, grows from zero. Since these axes lie at $45^\circ$ to the transverse coordinate directions, the common oscillation amplitude $A$ is
\begin{equation}
A=\sqrt{\frac{(\sqrt2\,a)^2+b^2}{2}}
=\sqrt{a^2+\frac{b^2}{2}},
\qquad
b=\sqrt{2(A^2-a^2)}.
\end{equation}
Thus $A=a$ gives the planar path, while $A=\sqrt2\,a$ gives the circular helix.

To preserve the crossing offsets, the phase shift $\alpha$ must satisfy $A\cos\alpha=a$. For a fixed crossing offset $a$, choosing the phase shift $\alpha$ therefore fixes the amplitude $A$, so either quantity can serve as the parameter for the interpolation. Adding equal and opposite shifts to the planar phases therefore gives
\begin{equation}
\begin{aligned}
\alpha&=\arccos(a/A),\\
\Phi_{ij}&=(1-c_{ij})\frac{\pi}{2}\pm\alpha
\quad [2\pi],\\
\Phi_{ij'}&=(1-c_{ij'})\frac{\pi}{2}\mp\alpha
\quad [2\pi].
\end{aligned}
\end{equation}
The signs are paired, and $\alpha$ varies from zero for the planar trajectory to $\pi/4$ for the circular helix. Throughout this range, the phases satisfy $A\cos\Phi_{ij}=a c_{ij}$, maintaining the prescribed offsets at each crossing.

Viewed end-on, the two sign choices trace the same ellipse in opposite directions as the yarn advances. They therefore give a right-handed and a left-handed trajectory, both passing through the same prescribed crossing offsets. At $\alpha=0$, the two choices coincide in the planar path. For nonzero $\alpha$, either handedness can be chosen independently for each yarn, allowing right- and left-handed trajectories to be combined within the same weave.

With the lag $\Phi_{ij'}-\Phi_{ij}=-\pi/2$ of Eq.~(\ref{eq_helix}), every yarn is right-handed, and all yarns are equivalent under the symmetries of the weave. The opposite lag, $\Phi_{ij'}-\Phi_{ij}=+\pi/2$, gives the mirror-image solution with every yarn left-handed, and since the lag is chosen independently for each yarn, the two may also be mixed. For the $2\times2\times2$ unit cell, whose twelve yarns each take one of the two lags, this gives $2^{12}=4096$ assignments, which reduce to $144$ distinct weaves under the rotations and translations that leave the weave invariant, and to $76$ when a weave and its mirror image are identified.

The renders in Figs.~\ref{fig_2Dweave}--\ref{fig_helix} were generated with a custom web-based visualizer~\cite{weaveviewer}, which computes the yarn trajectories of Section~\ref{sec_traj} for a chosen crossing configuration, trajectory family and number of unit cells. 
Readers are encouraged to interact with the visualizer to explore the different topologies and geometries of volumetric plain weaves. 
Claude Fable 5.1~\cite{claudefable} assisted with programming the visualizer.

 \subsection{Yarn intersections}\label{sec_intersection} 

We have established a method for generating the trajectories of the yarns of a volumetric plain weave directly from its crossing pattern.
The method ensures that the yarns pass through their prescribed positions at the lattice vertices. Between vertices, however, it is possible for centrelines to intersect, which could alter the intended topology and complicate fabrication. 
In this section, we determine the values of the offset $a$ for which the yarn centrelines do not intersect. 
We consider the weft, of trajectory $\bm{T_x}(t_x)$, and the warp, of trajectory $\bm{T_y}(t_y)$, of one vertex, with $t_x$ and $t_y$ the parameters of the two yarns along $\bm{\hat{x}}$ and $\bm{\hat{y}}$.
The analysis applies equally to the weft--wup and warp--wup pairs. 
We look for intersections in the neighbourhood of the vertex, at $t_x=t_y=0$, within $-\lambda/4\leq t_x,t_y\leq\lambda/4$, and we restrict the offset to $a\leq\lambda/4$, the value at which parallel yarns on neighbouring lattice lines, $\lambda/2$ apart and each offset by $a$ toward the other, come into contact. 

An intersection occurs when
\begin{equation}
\bm{T_x}(t_x^*)=\bm{T_y}(t_y^*),
\label{eq_dxyz}
\end{equation}
with $t_x^*$ and $t_y^*$ the parameters at intersection. Writing the coordinate differences as $\Delta x$, $\Delta y$, and $\Delta z$, this requires all three to vanish. We examine these conditions geometrically, first for the in-phase trajectories and then for the circular helices.

\subsubsection{In-phase} For the in-phase solution, each yarn's transverse coordinates follow the same cosine, multiplied by the prescribed crossing signs. These coordinates therefore remain proportional throughout the trajectory, confining the weft and warp to the planes
\[
y_\bot y=z_\bot z,
\qquad
x_\bot x=-z_\bot z,
\]
respectively. Any intersection must lie on the line shared by these planes. Eliminating $z$ between their equations gives $y_\bot y=-x_\bot x$. Since $x_\bot,y_\bot=\pm1$, this requires $|x|=|y|$. At an intersection, the weft and warp occupy the same point. Since $x=t_x$ for the weft and $y=t_y$ for the warp, the condition $|x|=|y|$ becomes $|t_x|=|t_y|$, or equivalently $t_x=\pm t_y$. Since cosine is an even function, this gives equal cosine factors,
\[
\cos\left(\frac{2\pi t_x}{\lambda}\right)
=
\cos\left(\frac{2\pi t_y}{\lambda}\right).
\]
The crossing matrix assigns opposite $z$-offsets to the weft and warp, $+a z_\bot$ and $-a z_\bot$. With equal cosine factors, their heights are therefore equal and opposite:
\begin{equation}
z_{\mathrm{weft}}
=a z_\bot\cos\left(\frac{2\pi t_x}{\lambda}\right),
\qquad
z_{\mathrm{warp}}
=-a z_\bot\cos\left(\frac{2\pi t_y}{\lambda}\right).
\label{eq_inter_inphase}
\end{equation}
These heights can coincide only at $z=0$. The plane equations then require $x=y=0$, leaving the origin as the only possible intersection. However, at $t_x=t_y=0$, the yarns have heights $+a z_\bot$ and $-a z_\bot$, which are nonzero for $a>0$. Thus, the in-phase trajectories have no local centreline intersections for any $a>0$, regardless of the crossing configuration. This local intersection condition therefore imposes no upper bound on the amplitude.

\subsubsection{Out-of-phase.} Consider the right-handed weft and warp at a $C_{\overline A}=C_{(-1,-1,-1)}$ vertex, following the circular helices of Section~\ref{sec_traj}. Viewed towards the vertex from the positive $x$- and $y$-directions, with $z$ upwards in both views, the yarns trace circular orbits of radius $A=\sqrt2\,a$. The $z$ components of the circular orbits differ in phase by $\pi$ ($\Phi_{xz}-\Phi_{yz}=\pi$), as prescribed by the phase relations in Eq.~(\ref{eq_helix}). Recall that this relation is fixed by the crossing configuration $C_{\overline A}$ and our choice of right-handed trajectories. At equal axial parameters, their projected positions therefore lie at opposite ends of a diameter.

Whereas the in-phase trajectories lie in fixed planes that meet along a line, the helical trajectories lie on perpendicular cylinders of radius $A$, described by
\[
y^2+z^2=A^2,\qquad x^2+z^2=A^2.
\]
Any intersection must satisfy both cylinder equations, giving $x^2=y^2$. The cylinders therefore meet along two ellipses, lying in the diagonal planes $y=x$ and $y=-x$. We examine these two planes in turn to determine whether the yarns intersect.

On $y=x$, the axial parameters are equal. The half-turn phase difference places the yarns at equal and opposite heights, so they can meet only at $z=0$. Their prescribed starting positions lie on diagonal radii of the projected circles. Tracing both orbits backwards through $45^\circ$ brings them to $z=0$. Since one full turn advances each yarn by $\lambda$ along its axis, this backward eighth-turn gives the candidate axial coordinate
\[
t_x^*=t_y^*=t^*=-\frac{\lambda}{8}.
\]
Within the vertex neighbourhood $-\lambda/4\leq t_x,t_y\leq\lambda/4$, this is the only candidate on $y=x$. At this point, the weft and warp are respectively at
\[
\left(-\frac{\lambda}{8},-A,0\right),
\qquad
\left(-A,-\frac{\lambda}{8},0\right).
\]
As $A$ increases, these points coincide precisely when
\begin{equation}
A=\frac{\lambda}{8},\qquad a_c=\frac{\lambda}{8\sqrt2}=\frac{\sqrt2\,\lambda}{16}\approx0.088\lambda.
\end{equation}

On $y=-x$, equality of the heights requires $\cos(2\pi t_x/\lambda)=0$. Within the vertex neighbourhood, the heights therefore agree only at $t_x=-t_y=\pm\lambda/4$, where the weft's $x$- and $y$-coordinates have the same sign. These points lie outside the required plane. The pair thus avoids centreline intersections for $0<a<a_c$, with the first intersection occurring at $a=a_c$. Above this value, the centrelines have passed through each other, inverting the intended over--under order.

The same critical offset applies to other crossing configurations and handedness choices. These change the signs of the prescribed offsets and the direction of rotation, but any permitted interior intersection still occurs at $z=0$, one eighth of a turn from the diagonal starting positions. The axial distance and required orbit radius therefore remain $\lambda/8$, giving $a_c=\lambda/(8\sqrt2)$. The argument also applies to the weft--wup and warp--wup pairs by permuting the coordinate axes. Although these choices determine which pairs intersect, $0<a<a_c$ provides a sufficient safe range for all of them.

\subsubsection{Finite radius overlap.} Finally, we consider the case when the trajectory are made of yarn of finite radii $R$, and the associated condition to avoid overlap between the yarns. 
We characterise the interlacing of two yarns by three quantities, here again described for a weft-warp pair, but generalizable to any other pair. 
\begin{itemize}
    \item The interlacing point $s$, which quantifies the distance from the vertex along each overlapping position, viewed from their common perpendicular direction. For a warp-weft pair, the interlacing point $s=|t_y^s|=|t_x^s|$ is defined by $\Delta x (t_x^s,t_y^s)=\Delta y(t_x^s,t_y^s)=0$ (but not necessarily $\Delta z(t_x^s,t_y^s)=0$).
    \item The separation $\Delta z^s=|\Delta z(t_x^s,t_y^s)|$ at the interlacing point, which quantifies how far apart the yarns are at their over-under crossing, it is positive while the interlacing is as prescribed and changes sign when the yarns pass through each other.
    \item The closest approach $\Delta_{\min}$, the numerically computed minimum distance between the two yarns. This value is bounded by $\Delta z^s$, but is possibly smaller when trajectories have a strong $z$ component near the interlacing point. The necessary condition for no yarn overlap is then $\Delta_{\min}>2R$.
\end{itemize}

For the in-phase case, we directly derive the interlacing and separation from Eq.~\ref{eq_inter_inphase}:
\begin{equation}
    s=a\cos\frac{2\pi s}{\lambda}, \quad \Delta z^s=2a\cos\frac{2\pi s}{\lambda}. \label{eq_pair_in}
\end{equation}
To derive analytical solutions, we take the limit $a/\lambda\ll1$ to first order. We obtain simple expressions: $s=a$ and $\Delta z^s=\Delta_{\min}=2a$.
The non-overlap condition is simply written as $a\geq R$. 
We also notice that, as $s=a$, the yarn in a volumetric weave does not cross at the vertex but along the diagonals emerging from it. 
It is in contrast with the 2D case where $s=0$; sections of a volumetric weave therefore differ slightly from the planar case (see Fig.~\ref{fig_3Dunit}d).

For the out-of-plane case, we focus on two situations: when the yarns intersect at $a_c=\lambda/(8\sqrt2)$, and when they drift apart without intersecting for any $a<\lambda/4$. 
In both cases, all the yarns are taken right-handed, but they differ by their crossing configurations, taken as $C_{(\epsilon,\epsilon,\epsilon)}$, with $\epsilon=\pm1$. 
The case $\epsilon=-1$ is the one considered earlier, with intersecting yarns, while in the second one $\epsilon=1$, yarn do not intersect.
For these configurations, imposing $\Delta x=\Delta y=0$ on the helical trajectories of Section~\ref{sec_traj} gives the interlacing point and the corresponding height difference:
\begin{equation}
    s=a\Big(\cos\frac{2\pi s}{\lambda}-\epsilon\sin\frac{2\pi s}{\lambda}\Big), \quad \Delta z^s=2a\Big(\epsilon\cos\frac{2\pi s}{\lambda}+\sin\frac{2\pi s}{\lambda}\Big). \label{eq_pair_out}
\end{equation}
These equations have no closed-form solution, and we solve them numerically.

\begin{figure}[ht]
\centering
\includegraphics[width=\textwidth]{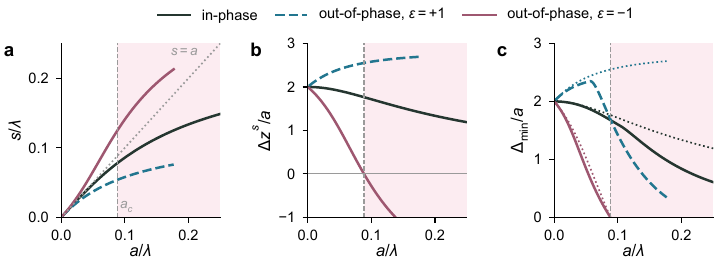}
\caption{\textbf{Weft--warp pair of a vertex as a function of the offset $a$}. The in-phase pair and the right-handed out-of-phase pairs at a non-intersecting vertex ($\epsilon=+1$) and at an intersecting vertex ($\epsilon=-1$). \textbf{a} Distance $s$ of the interlacing point from the vertex and \textbf{b} separation $\Delta z^s$ at the interlacing point, positive while the interlacing is as prescribed, solved from Eqs.~(\ref{eq_pair_in}) and (\ref{eq_pair_out}). \textbf{c} Closest approach $\Delta_{\min}$ between the two centrelines, computed numerically (solid), compared with $\Delta z^s$ (dotted). The in-phase curves end at $a=\lambda/4$, where parallel yarns on neighbouring lattice lines touch, the out-of-phase curves at $a=\lambda/(4\sqrt2)$, where the cylinders of radius $\sqrt2\,a$ about neighbouring lattice lines touch. Beyond $a_c$ (shaded), the pair at the intersecting vertex is inverted.}\label{fig_interlacing}
\end{figure}

The resulting variations of the interlacing point $s$, distance $\Delta z^s$ and closest approach $\Delta_{\min}$ are plotted against the offset $a$ in Fig.~\ref{fig_interlacing}. 
These curves may serve as a design basis to evaluate the admissible yarn radii $R$ such that $2R<\Delta_{\min}$. 
We recover that $\Delta z^s$ vanishes at $a=a_c$ and changes sign beyond for the helical intersecting case $\epsilon=-1$, but stays positive for the in-phase and $\epsilon=+1$ cases (see Fig.~\ref{fig_interlacing}b). 
For small offsets, the separation is a good approximation for the closest approach ($\Delta z^s \approx \Delta_{\min}$), but in particular for the $\epsilon=1$ and in-phase cases, the two curves drift apart with increasing $a$, imposing stricter conditions on the acceptable radii $R$. 

\section{Crafting of volumetric weaves}
The trajectories developed above provide a direct basis for fabricating volumetric weaves. We first use additive manufacturing to realize the prescribed geometry, then manually construct the model from compliant yarns.

\subsection{Additive manufacturing crafting}

\begin{figure}[ht!]
\centering
\includegraphics[width=\textwidth]{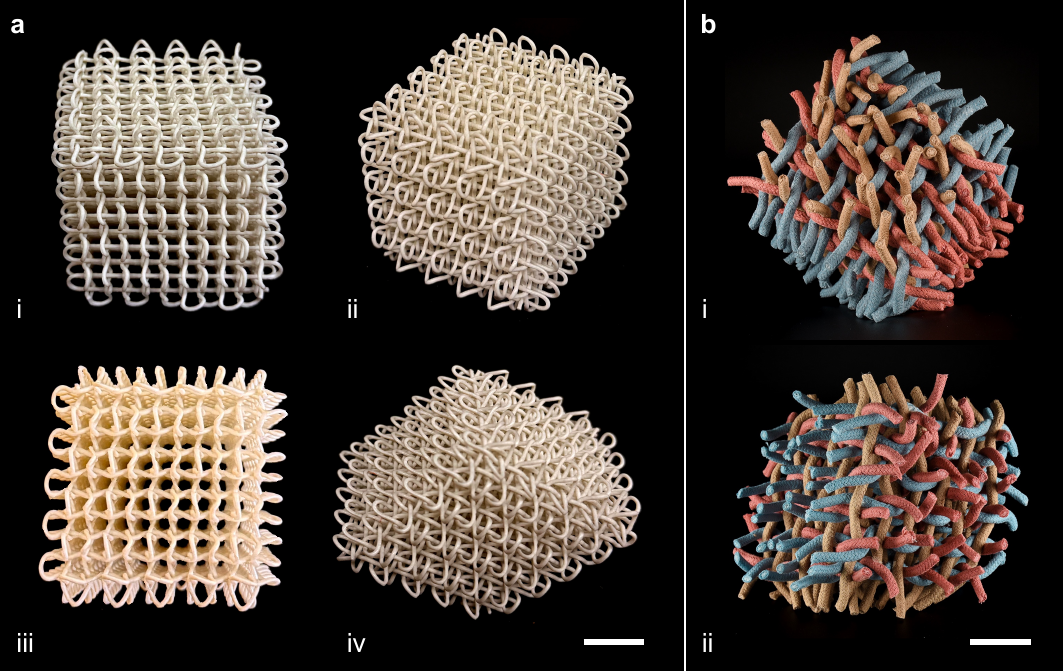}
\caption{\textbf{Realized models of a volumetric plain weave.} \textbf{a} 3D-printed model with in-phase yarn trajectories and five units (ten crossings) in each of the weft, warp and wup directions: i--iv, different views of the same specimen. Scale bar, 10~mm. \textbf{b} Two views of the model manually assembled from cotton cord in three colours. Scale bar, 25~mm.}\label{fig:print}
\end{figure}

In Fig.~\ref{fig:print}a, we show different views of one 3D-printed volumetric weave with in-phase yarn trajectories. The specimen follows Eq.~(\ref{eq_inphase}) with the configuration $C_A$ at the origin, a wavelength $\lambda=10$~mm, an offset $a=0.7$~mm and a yarn radius $R=0.5$~mm, and comprises five unit cells along each direction, so that each yarn passes through ten crossings. To fabricate this model, we implemented the parametric curves in Python and visualized them using the Grasshopper+Rhino3D CAD package. With this parametrization, the closest distance between the centrelines of two crossing yarns is $s_{\min}=1.25$~mm (Fig.~\ref{fig_interlacing}c), which exceeds the yarn diameter. 

At the boundaries, each loose yarn was systematically connected to the next yarn, making each of the weft, warp and wup families a single continuous yarn. The curves were then given the radius $R$ to create curved cylindrical rods and surface meshed. The model was printed on a Stratasys J35 using VeroUltraWhite as the print material and a sacrificial support material to preserve the topology. The support material was dissolved in water over the course of 24 hours, and the printed model was dried for a further 24 hours.

We observe a stiff response along the principal yarn axes and compliance between them, as is characteristic of plain weaves. The specimen relaxes under gravity in a direction at $45^\circ$ to the principal yarn axes (see Fig.~\ref{fig:print}a\,iv). Further mechanical characterization will be explored in future work.

\subsection{Manual crafting}

In Fig.~\ref{fig:print}b, we show two views of a volumetric weave manually assembled from cotton cord. To make this textile, we used the ideal geometry as a guide for assembly. In conventional weaving, a loom maintains tension that helps regulate yarn spacing and the articulation of yarns around one another at each crossing. Here, no such loom was available, so we pre-shaped the yarns to guide their local curvature and spacing during assembly while retaining sufficient pliability for weaving. Cotton piping cord of 6~mm diameter in three colours (Amazon.com) was shaped into planar sinusoids using a 3D-printed mold and a commercial latex-paint conditioner (Flood Floetrol; Amazon.com) as a temporary shape-setting medium. The shaped weft and warp yarns were first assembled into two-dimensional layers, keeping the slant of the crossings consistent within each layer, following the in-phase geometry developed in Section~\ref{sec_traj}. All 36 wup yarns were introduced, and successive layers were woven into them layer by layer and row by row. Each layer was progressively peeled back to give access to the crossings as the wup yarns were threaded through the appropriate positions. Assembly was complete when the final two-dimensional layer had been incorporated. Boundary vertices were glued during construction, and selected interior vertices were bonded to assist assembly; most interior vertices remained unbonded.

Despite efforts to maintain regular spacing and orthogonal directions, the textile sheared both within and out of the layer planes as it relaxed, giving the specimen rhomboidal faces (see Fig.~\ref{fig:print}b\,ii). The observed shear may reflect interactions between yarn elasticity, friction, geometry, and topology. Further work is needed to distinguish their contributions and assess the role of assembly history.

\section{Discussion and Conclusion}

We have described volumetric plain weaves by a crossing matrix at each vertex of a cubic lattice, which records the offsets of the three yarns passing through it. The cancellation of the offsets leaves eight configurations related by rotations, so that the alternating volumetric weave has a single topology, as its planar counterpart. The topology and associated offset sequences then guide the geometry of the yarn trajectories. The offsets fix the amplitude and phases of sinusoidal yarn trajectories and select two solutions, planar zigzags and helices, which belong to one family of elliptical helices. We showed above that in-phase trajectories have no local centreline intersections for any $a>0$, whereas circular helices can intersect at $a_c=\lambda/(8\sqrt2)$. For the helical family, $0<a<a_c$ therefore provides a sufficient range for preserving the local crossing topology. Within these bounds, the weaves were fabricated by additive manufacturing. Manual assembly of the in-phase construction from compliant yarns provides a complementary exploration. The crossing pattern determines how the yarns are placed during assembly, while the observed shear upon relaxation raises an open question about the interaction of geometry, topology and yarn mechanics.

This work suggests several directions for continued investigation. Patterns in which a yarn passes several consecutive crossings on the same side, the volumetric counterparts of twill and satin, require trajectories of longer period with several harmonics, whose coefficients are again fixed by the offsets at the vertices; the crossing matrix and the analysis of intersections carry over. The mechanical characterisation of the weaves is the next step: the soft modes of yarns bending and sliding between vertices, the elasticity as a function of the three lengths $\lambda$, $a$ and $R$ and of the trajectory family, and the response to defects, from an inverted crossing to a missing yarn, up to fracture. Manual crafting proceeds layer by layer and exposes the operations that a volumetric loom would automate: a periodic sequence of displacements of the three yarn families, comparable to the opening of a shed, fed by continuous yarns as in the printed specimens. Finally, the trajectories provide centrelines for simulation and fabrication at any scale, for composites reinforced through their thickness \cite{li2022review}, woven metamaterial lattices \cite{carton2026woven} and molecular weaving \cite{liu2016weaving}.

\textbf{Acknowledgements}
S.P. acknowledges the financial support provided by JSPS KAKENHI Grant-in-Aid for Early-Career Scientists No. 25K17363. L.C.D. acknowledges support from the Chancellor's Fellowship at the University of California, Santa Barbara. T.C. acknowledges support from the U.S. National Science Foundation through the DMREF program under Grant No.~2523081.

\textbf{Disclosure of Interests}
The authors have no competing interests to declare that are relevant to the content of this article.

\bibliographystyle{splncs04}
\bibliography{references}

\end{document}